\documentclass[aps,twocolumn,floats,prl,footinbib]{revtex4}
\usepackage{graphics,graphicx,epsfig}
\usepackage{amssymb,color}
\usepackage{amsmath,textgreek,bm}
\usepackage{epsf,epstopdf,wrapfig}
\usepackage{xcolor}
\usepackage{tikz}
\def\debug{1}
\newcommand{\Fnotes}[1]{\ifnum\debug=1{\color{purple} [FM: #1]}\fi}

\newcommand{\Cnotes}[1]{\ifnum\debug=1{\color{teal} [FM: #1]}\fi}

\newcommand{\LDC}[1]{\ifnum\debug=1{\color{red} [LDC: #1]}\fi}

\newcommand{\sbold}{\boldsymbol{s}}

\newcommand{\ON}{\mathcal{O} \left (N^{-1} \right) }

\date{\today}

\begin{document}
\title{Extended mean--field theories for networks of real neurons}

\author{Luca Di Carlo,$^a$ Francesca Mignacco,$^{a,b}$ Christopher W. Lynn,$^{a,b,c}$ and William Bialek$^{a,b}$}
\affiliation{$^a$Joseph Henry Laboratories of Physics and Lewis--Sigler Institute, Princeton University, Princeton NJ 08544 USA}
\affiliation{$^b$Initiative for the Theoretical Sciences, The Graduate Center, City University of New York, 365 Fifth Ave, New York NY 10016 USA}
\affiliation{$^c$Department of Physics, Quantitative Biology Institute, and Wu Tsai Institute, Yale University, New Haven CT 06510 USA}

\date{\today}
\begin{abstract}
If the behavior of a system with many degrees of freedom can be captured by a small number of collective variables, then plausibly there is an underlying mean--field theory.  We show that simple versions of this idea fail to describe the patterns of activity in networks of real neurons.  An extended mean--field theory that matches the distribution of collective variables is at least consistent, though shows signs that these networks are poised near a critical point, in agreement with other observations.  These results suggest a path to analysis of emerging data on ever larger numbers of neurons.
\end{abstract}

\maketitle

Throughout much of the brain, the electrical activity of single neurons is effectively binary, and thus equivalent to  Ising spins.  When we look at many neurons 
%simultaneously we see patterns of activity which we can think of as configurations of $N$ Ising spins, and 
we would like to know the distribution out of which  spin configurations are drawn.  There is a long history of using ideas from statistical physics to approach this problem \cite{Meshulam+Bialek_2024}.

Important landmarks in the statistical physics approach to neural networks are the Hopfield model \cite{hopfield1982} and the Boltzmann machine \cite{ackley+al_85}.  In the Hopfield model the dynamics have a Lyapunov or energy function, and local energy minima are stable configurations that we can identify as stored memories or   as the results of computations \cite{hopfield+tank1985,hopfield+tank1986}.  In the Boltzmann machine the states of ``visible'' neurons represent signals in the outside world, ``hidden'' neurons represent inferred causes or features of these variables, and the distribution over network states is meant to simulate the distribution over signals.  Both models can be described by pairwise interactions among the Ising spins, with learning rules that adjust these interactions to reach the desired behaviors.

There is a path that leads directly from experimental data to pairwise Ising models that are like the Hopfield model or Boltzmann machine \cite{Meshulam+Bialek_2024,schneidman+al_2006}.  Concretely, we write the state of each neuron as $s_n$, with $s_n = +1$ when the cell is active and $s_n = -1$ when the cell is silent; the state of the network is $\sbold \equiv \{s_1,\, s_2,\, \cdots ,\, s_N\}$.  We are looking for the probability distribution $P(\sbold)$ and insist that it match the observed means and correlations,
\begin{eqnarray}
\langle s_n\rangle_P &=& \langle s_n \rangle_{\rm exp} \label{pair1}\\
\langle s_n s_m \rangle_P &=& \langle s_n s_m \rangle_{\rm exp} ,\label{pair2}
\end{eqnarray}
where $\langle \cdots \rangle_P$ denotes an average over the distribution $P$ and $\langle \cdots \rangle_{\rm exp}$ denotes an average over the data.  There are infinitely many distributions that satisfy these constraints, but among these one is special because it generates states that are as random as possible and hence introduces no knowledge or structure beyond that required to match the expectation values in Eqs (\ref{pair1}, \ref{pair2}); this is the maximum entropy distribution \cite{jaynes1957information, jaynes1982rationale} 
%and has the form of a Boltzmann distribution
\begin{eqnarray} 
P_{\rm pairs}(\sbold) &=& \frac{1}{Z_{\rm pairs}} \exp\left[ - E_{\rm pairs} (\sbold )\right] ,\\
    Z_{\rm pairs}  &=& \sum_{\sbold} \exp\left[ - E_{\rm pairs} (\sbold )\right]  ,\\
 \label{Zpairs}
% \end{eqnarray}
% with the energy function
% \begin{equation}
 E_{\rm pairs} (\sbold ) &=&   -  \sum_n h_n s_n - {1\over 2}\sum_{n,m = 1}^N  s_n J_{nm} s_m  .
%\end{equation}
 \end{eqnarray}
We recognize this as the Boltzmann distribution for a particular Ising spin glass in which the fields $\{h_n\}$ and couplings $\{J_{nm}\}$ are determined by the data through Eqs (\ref{pair1}, \ref{pair2}); once we solve these equations all further predictions are parameter free.  These quantitative predictions have been very successful
% for example predicting all $\sim 10^5$ triplet correlations among $N\sim 100$ neurons in the mouse hippocampus, and other collective behaviors 
in several different systems \cite{Meshulam+Bialek_2024}.

The pairwise maximum entropy model  begins by taking $\sim N^2$ expectation values from the data.  But current experimental methods allow for dramatic increases in  $N$ without proportional increases in the duration $T$ of experiments; eventually the measured matrix of pairwise correlations fails to be of full rank, limited by $T$ and not $N$.   It is acceptable to ask for experimental estimates of $M\propto N$ expectation values, and then construct maximum entropy models that are consistent with these measurements.  As an example, in an Ising model with local interactions we can recover the correct Boltzmann distribution by asking for the maximum entropy model consistent with near-neighbor correlations, and there are ${\cal O}(N)$ of these even without translation invariance.  This reconstruction from local correlations also works in flocks of birds \cite{bialek2012statistical,cavagna+al_2015}.  How do we choose ${\cal O}(N)$ constraints  in the absence of guidance from symmetries or locality?
The literature on patterns of activity in populations of neurons offers several intuitions.
% that may be of help in choosing constraints.  

{\em Population activity.} It has been suggested that collective effects in networks of neurons can be captured by keeping track of the summed activity \cite{okun+al_2012,tkacik+al_2013},
\begin{equation}
\phi = \sum_{n=1}^N s_n .
\label{def:phi}
\end{equation}
The maximum entropy model that is consistent with the mean and variance of $\phi$ again has the form of a Boltzmann distribution, now with the energy function
\begin{equation} 
E_{\rm pop}(\sbold) = - h  \phi  - \frac{\lambda}{2N}  \phi^2 ,
\label{Epop}
\end{equation}
which we recognize as a mean-field ferromagnet \cite{parisi1988statistical,sethna2021statistical,kivelson+al_24}.  We recall that in solving this model we rewrite it (exactly) as spins interacting with an auxiliary field.  This connects to the intuition that the high dimensional dynamics of a network is controlled by a small number of ``latent fields'' to which the cells are responding independently \cite{sahani_99,whiteway+butts_2019};  here this latent field is  an  emergent property   rather than  externally imposed.  At large $N$ fluctuations in the latent field are suppressed, and we refer to this as the naive mean-field approximation. 

{\em Means and projections.} Focusing on the summed activity we lose the identities of the  neurons.  An alternative is to ask for a model that matches the mean activity of each individual cell, but tries to capture collective effects by looking only along one or a few projections
\begin{equation}
\varphi_\alpha = \sum_{n=1}^N W_{\alpha n} s_n .
\end{equation}
The maximum entropy model that matches all the mean activities $\langle s_n\rangle$ and the covariance matrix $\langle \varphi_\alpha \varphi_\beta\rangle$ is a Boltzmann distribution with energy
\begin{equation}
E_{\rm proj}(\sbold ) = -\sum_{n=1}^N h_n s_n -{1\over {2N}}\sum_{\alpha,\beta = 1}^K \varphi_\alpha \Lambda_{\alpha\beta}\varphi_\beta ,
\label{Eproj}
\end{equation}
where we keep track of $K$ projections.  This connects to the intuition that the high dimensional dynamics of a network are dominated by a low dimensional manifold \cite{Cunningham2014,gallego2017neural,Edward2021}.  Solving the model again introduces auxiliary fields, one conjugate to each projection $\varphi_\alpha$, so this is a (slightly) generalized mean-field theory \cite{Cocco2011}.  For $K \ll N$ fluctuations in these fields are suppressed at large $N$, which again we refer to as naive mean-field.  If $K\propto N$ there should be a more sophisticated mean-field theory analogous to the Hopfield network near saturation \cite{amit+al_1987}.

{\em Distributions of projections.} The emphasis on the variance of activity along particular projections may be too limiting.  An alternative is to match the mean activity of each neuron and the distribution of activity along a limited set of directions in the high dimensional space of neural activity.  For simplicity we consider here the case of just one projection $\varphi$.  The maximum entropy model now is a Boltzmann distribution with energy  
\begin{equation}
E_{\rm dist}(\sbold ) = -\sum_{n=1}^N h_n s_n + N U (\varphi ) .
\label{Edist}
\end{equation}
As usual the fields $\{h_n\}$ are adjusted to match the measured mean activity of each neuron and the effective potential $U (\varphi )$ is adjusted to match the observed distribution $P_{\rm exp} (\varphi )$; the factor $N$ keeps the potential  of order one.  This is equivalent to a model with a single latent field acting on all the neurons, with an arbitrary distribution.  It also connects to models for dense associative memory, or ``modern Hopfield'' networks \cite{krotov+hopfield_2016}.  Our major result  is the construction of a generalized mean-field theory for this class of models and the demonstration that real networks are in the regime where this approximation is valid.  In contrast, we will see that simpler mean-field theories fail in their naive form, and if generalized they provide a bad description of higher order structure in the data.  A fuller account   will be given elsewhere \cite{dicarlo+al_2025}.

Calculating the partition function of the population activity model [Eq~(\ref{Epop})] is a standard exercise   \cite{parisi1988statistical,sethna2021statistical,kivelson+al_24}.  The partition function can be rewritten exactly as 
%integral over the auxiliary field $\psi$,
\begin{eqnarray}
Z_{\rm pop} (h,\lambda )&=&\sqrt{N\over{2\pi\lambda}} 2^N 
 \int d\psi\, e^{-N f(\psi )},
 \label{Zpop}\\
 f(\psi ) &=& \frac{1}{2\lambda}\psi^2 - \ln\cosh(h + \psi) .
\label{fpop}
\end{eqnarray}
At large $N$ the integral in Eq (\ref{Zpop}) should be dominated by the the auxiliary field $\psi = \psi_*$ that minimizes the free energy $f(\psi )$, and this leads to the naive mean-field approximation
\begin{equation}
\ln Z(h,\lambda )=-N f(\psi_* ) + N\ln 2 - \frac{1}{2}\ln[2\pi\lambda f''(\psi_*)]  + \cdots ,
\label{lnZ_pop}
\end{equation}
where $\cdots$ are terms $\sim 1/N$.   In this approximation 
\begin{eqnarray}
\langle \phi \rangle &\equiv & \mu = \tanh (h + \lambda \mu ) +  \ON,\\
{1\over N} \langle (\delta \phi )^2 \rangle &\equiv& \chi = \frac{ 1- \mu^2}{ 1 - \lambda(1- \mu^2)} + \mathcal O \left ( N^{-1} \right )  .
\label{chi_def1}
\end{eqnarray}
These self--consistent equations can be combined to show that there is a maximum $\chi$ at fixed $\mu$,
\begin{equation}
    \chi_{\mathrm{max}}(\mu) = \frac{ \mu(1-\mu^2)}{\mu   - {\mathrm{atanh}(\mu)} (1-\mu^2)  } \ .
    \label{eq:nmf_chimax}
\end{equation}

\begin{figure}
\centering
\includegraphics[width=0.9\linewidth]{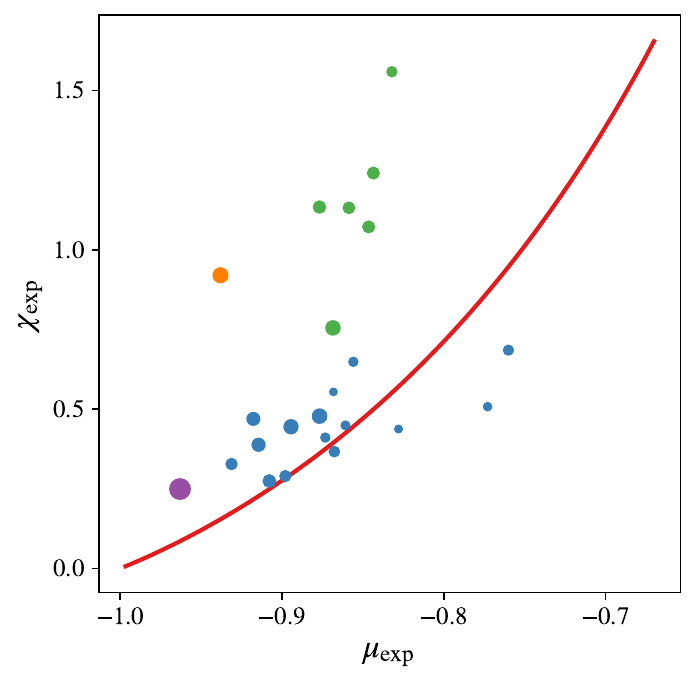}
\caption{Susceptibility and magnetization in networks of real neurons (points) compared with the bound from naive mean-field theory (red line).  Data from multiple different experiments: $N=160$ neurons in the retina  (orange), recorded with an electrode array \cite{tkavcik2014searching}; $N=1416$ neurons in the mouse hippocampus (purple), recorded with a fluorescent calcium indicator \cite{gauthier2018dedicated,meshulam2019RG}; $N = 60-190$ neurons from single brain regions in the mouse (blue)   and $N= 900 - 1500$ neurons across multiple regions (green), recorded with Neuropixels 2.0 \cite{AllenData}. \label{chi_mu_data}}
\end{figure}

In Figure \ref{chi_mu_data} we show $\chi$ vs $\mu$ computed from data on several systems:
$N=160$ neurons at the output of the vertebrate retina, the  cells that send information from eye to brain \cite{tkavcik2014searching}; groups of $N \sim 60-190$ neurons in each of several regions of the mouse brain, or $N\sim 900-1500$ neurons across multiple regions  \cite{AllenData}; and $N\sim 1500$ neurons in the mouse hippocampus \cite{gauthier2018dedicated,meshulam2019RG}.    Measured values of  $\chi$ consistently exceed the bound from 
%naive mean--field theory, 
Eq (\ref{eq:nmf_chimax}).

\begin{figure}[b]
\centering
\includegraphics[width = 0.9\linewidth]{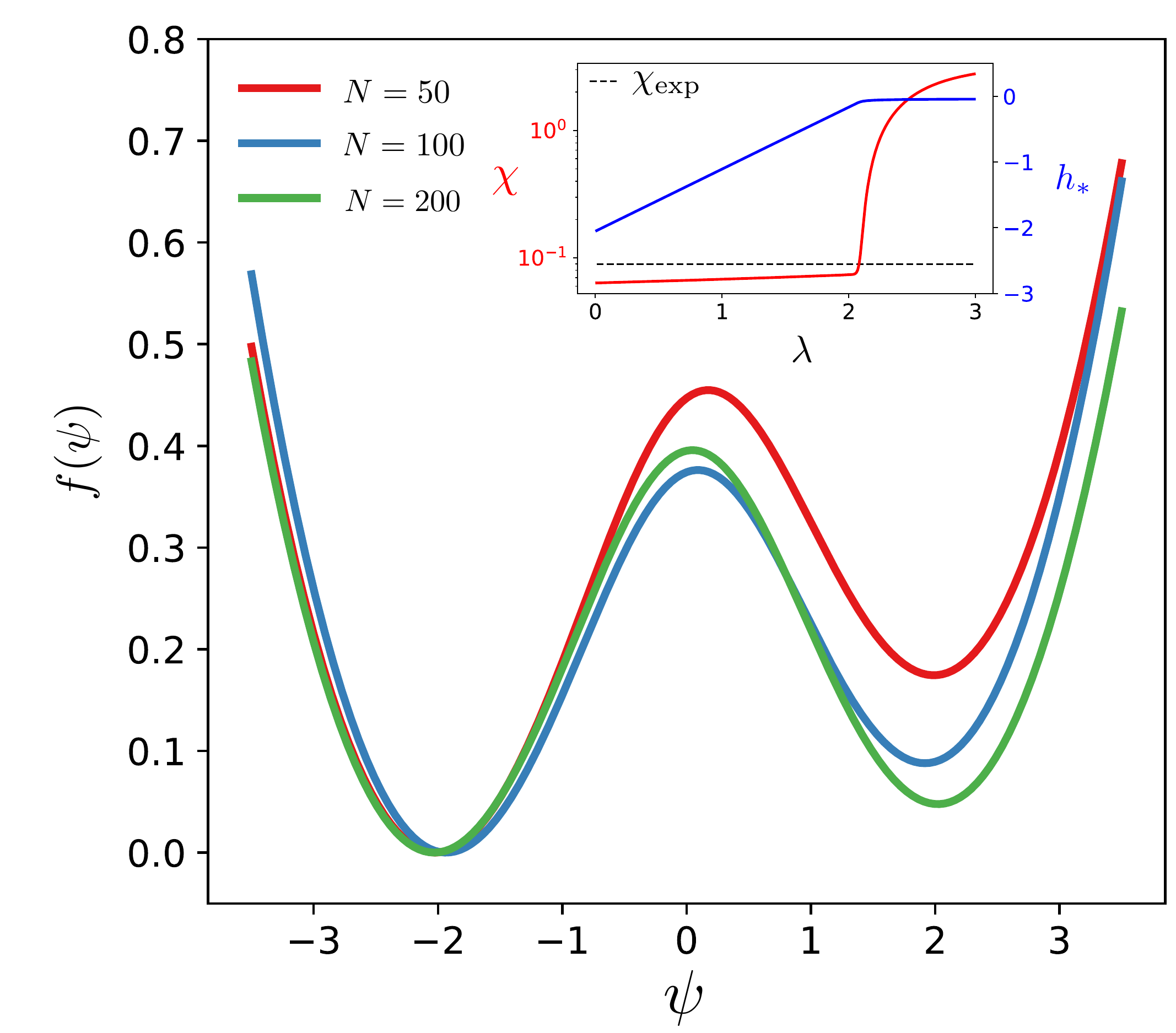}
\caption{Free energy per neuron $f(\psi )$ in the model that matches the mean and variance of summed population activity, Eq (\ref{fpop}).  Parameters $h$ and $\lambda$ inferred from experiments  on a network of neurons in the mouse hippocampus \cite{gauthier2018dedicated,meshulam2019RG}.  As we consider larger populations (increasing $N$), the two local minima become more nearly degenerate, signaling the proximity of a first order phase transition. The inset shows the trajectory $h_*(\lambda)$ (blue) with constant magnetization  and the corresponding susceptibility $\chi$ (red) for a network of $N=50$ neurons; the intersection of $\chi$ with  $\chi_{\rm exp}$ (dashed line) determines the exact solution of the maximum entropy problem. \label{fig:fpop}}
\end{figure}

The conclusion from Fig \ref{chi_mu_data} is that real neurons are not consistent with naive mean-field theory.  But what happens to the maximum entropy problem?  That is, can we still find values of $h$ and $\lambda$ in Eq (\ref{Epop}) such that the predicted $\mu$ and $\chi$ agree with measured values? The answer is yes, but it must be done very carefully \cite{dicarlo+al_2025}.  We can use the exact integral representation of $Z_{\rm pop}$ in Eq (\ref{Zpop}) to construct, numerically, a trajectory $h_*(\lambda )$ along which the predicted magnetization $\mu=\mu_{\rm exp}$ (Fig \ref{fig:fpop}, inset).  This trajectory begins at $h_*(\lambda = 0) = \mathrm{atanh}(\mu_{\rm exp})$ and $h_*$ moves toward $h_* = 0$ as $\lambda$ increases.  The trajectory stalls, with $dh_*/d\lambda$ becoming almost zero at a critical $\lambda$, at which point the susceptibility $\chi$  rises rapidly; in the examples from Fig \ref{chi_mu_data}, $\chi_{\rm exp}$ intersects this steep rise, so that $\lambda$ is determined very precisely by the data.

Having found $h$ and $\lambda$ to match the mean and variance of the population activity $\phi$, we can look directly at the free energy $f(\psi )$ from Eq (\ref{fpop}); results are shown in Fig \ref{fig:fpop} for a population of neurons in the mouse hippocampus \cite{gauthier2018dedicated,meshulam2019RG}.  
There are two local minima,  very nearly degenerate.The population of hippocampal neurons analyzed here lives in a plane, and so we can change $N$ by expanding the radius of a circle \cite{Meshulam2021,Lynn2023}.  As $N$ increases the two minima of the free energy become more nearly degenerate, and indeed the free energy difference scales as $1/N$.  This  is a defining feature of a first order phase transition.   We emphasize that this is driven entirely by the experimental observations $\mu_{\rm exp}$ and $\chi_{\rm exp}$.

The fact that the data drive these minimally structured models toward a first order transition is provocative.  This result explains the failure of the naive mean-field approximation and is suggestive of a network balanced between macroscopically ordered patterns of activity.  Unfortunately,  while this model (by construction) matches the mean and variance of the summed population activity, the presence of two local minima in $f(\psi)$  predicts that the distribution of this summed activity will be bimodal, and this is qualitatively wrong.

The problems that we find in matching the mean and variance of population activity persist in models that match the mean activity of individual neurons and variances of projections, as in Eq (\ref{Eproj}), though the details depend on our choice of the projections.  If we choose a single projection with coefficients $W_{\alpha n}$ given by independent Gaussian random numbers, then the naive mean-field approximation works but the resulting model is not very interesting, since it has an entropy only slightly below that for completely independent neurons. 

To order possible choices of the projection coefficients it is useful to think about the correlation matrix
\begin{equation}
\tilde C_{nm} \equiv {{\langle s_n s_m\rangle - \langle s_n \rangle\langle s_m\rangle}
\over{\left[ (1-\langle s_n \rangle^2) (1-\langle s_m \rangle^2) \right]^{1/2}}} .
\end{equation}
This matrix has eigenvectors $\tilde {\mathbf W}$ and eigenvalues $\rho$,
\begin{equation}
\sum_{m=1}^N \tilde C_{nm}\tilde W_{\alpha m} = \rho_\alpha \tilde W_{\alpha n} .
\end{equation}
If we choose  projections with coefficients
\begin{equation}
W_{\alpha n} = {1\over\sqrt{1-\langle s_n \rangle^2}} \tilde W_{\alpha n},
\label{WWtilde}
\end{equation}
then following the same naive mean-field approximation that leads to Eq (\ref{lnZ_pop}) 
%shows that the model in Eq (\ref{Eproj}) has 
gives the entropy \cite{Cocco2011,dicarlo+al_2025}
\begin{equation}
S = S_0 - {1\over 2}\sum_\alpha [\rho_\alpha - 1 - \ln (\rho_\alpha )] ,
\end{equation}
where $S_0$ is the entropy of independent neurons.  Thus the most informative projections are those connected to the largest or smallest eigenvalues of the correlation matrix.  But if we choose  these optimized projections we are led back to the problem of degenerate local minima.
% that we found when constraining only the summed population activity.

To tame the multiple minima we  insist on matching both the mean activities of individual neurons and the {\em distribution} of activities along a projection, leading to  Eq (\ref{Edist}).  
%Solving this problem requires us to estimate the fields $\{h_n\}$ and also to reconstruct the function $U(\varphi )$.  To begin, the 
The partition function 
\begin{eqnarray}
Z_{\rm dist} &\equiv& \sum_{\sbold} 
\exp\left[ \sum_{n=1}^N h_n s_n - N U\left( \sum_{n=1}^N W_n s_n \right)\right] \\
%\end{equation}
%can be written as
%\begin{equation}
%Z_{\rm dist} 
&=& 2^N \int {{dz}\over{2\pi}} \int d\varphi\,\exp\left[ -F_{\rm dist}(\varphi ; z)\right] ,
\label{Zdist3}
\end{eqnarray}
where the free energy
\begin{equation}
F_{\rm dist}(\varphi ; z) = iz\varphi + N U(\varphi ) -\sum_{n=1}^N \ln\cosh(h_n + izW_n ) .
\end{equation}
%
%\begin{widetext}
%\begin{eqnarray}
%Z_{\rm dist} 
%%&\equiv& \sum_{\sbold} 
%%\exp\left[ \sum_{n=1}^N h_n s_n - N U\left( \sum_{n=1}^N W_n s_n \right)\right]\\
%&=& \sum_{\sbold} \int {{dz}\over{2\pi}} \int d\varphi\,
%\exp\left[ -  iz\left(\varphi - \sum_{n=1}^N W_n s_n\right) + \sum_{n=1}^N h_n s_n - N U\left(\varphi \right)\right]\\
%&=& 2^N \int {{dz}\over{2\pi}} \int d\varphi\,\exp\left[ -F_{\rm dist}(\varphi ; z)\right] ,
%\label{Zdist3}
%\end{eqnarray}
%\end{widetext}
%where the free energy
%\begin{equation}
%F_{\rm dist}(\varphi ; z) = iz\varphi + N U(\varphi ) -\sum_{n=1}^N \ln\cosh(h_n + izW_n ) .
%\end{equation}
As usual we search for a saddle point  $(\varphi_{\rm sp}, z_{\rm sp})$ that extremizes $F_{\rm dist}(\varphi ; z)$, and we find  
\begin{eqnarray}
    \varphi_{\rm s p} &=& \sum_{n=1}^N W_{n}  \tanh \left ( h_{\rm n} + i z_{\rm s p} W_{n} \right )  \\ 
    z_{\rm s p } &=& i N U^{\prime}(\varphi_{\rm s p}) ;
\end{eqnarray}
we will then approximate
\begin{equation}
\ln Z_{\rm dist} \approx -F_{\rm dist}(\varphi_{\rm sp}; z_{\rm sp}) + \cdots ,
\end{equation}
where $\cdots$ are of less importance at large $N$.
Notice that $E_{\rm dist}(\sbold )$ is invariant under the gauge transformation $U(\varphi ) \rightarrow U(\varphi ) - \varphi U^{\prime}(\varphi_{\rm s p})$ and $h_n \rightarrow h_n + NU^{\prime}(\varphi_{\rm s p})W_n$, so we can set $U^{\prime}(\varphi_{\rm s p}) = 0$.
%Notice that if we shift $U(\varphi ) \rightarrow U(\varphi ) - \varphi U^{\prime}(\varphi_{\rm s p})$ and $h_n \rightarrow h_n + U^{\prime}(\varphi_{\rm s p})W_n$ then the energy $E_{\rm dist}(\sbold )$ is unchanged.  We can exploit this gauge invariance to set $U^{\prime}(\varphi_{\rm s p}) = 0$.  
This fixes $z_{\rm zp} = 0$ and yields 
\begin{equation}
\langle s_n\rangle =  {{\partial Z_{\rm dist}}\over{\partial h_n}} = \tanh h_n ,
\end{equation}
as if the neurons responded independently to the fields
%.  Matching the mean activity of each neuron then gives
%Setting 
\begin{equation}
h_n = \mathrm{atanh}(\langle s_n \rangle_{\rm exp}).
\label{h_Edist}
\end{equation}
%thus guarantees that our model matches the mean activity of each neuron.

The distribution of activity along the projection is 
\begin{equation}
P(\varphi ) = {{2^N}\over{Z_{\rm dist}}}  \int {{dz}\over{2\pi}}\exp\left[ -F_{\rm dist}(\varphi ; z)\right] .
\end{equation}
The saddle point approximation is
\begin{equation}
P(\varphi ) \propto \exp\left[ -F_{\rm dist}(\varphi ; z_*(\varphi))\right] ,
\label{Pphi}
\end{equation}
where $z_*$ depends on $\varphi$ and is the solution of 
\begin{equation}
\sum_{\rm n=1}^N W_n \tanh\left[ h_n + iW_n z_*(\varphi) \right] = \varphi .
\label{zstar}
\end{equation}
Given weights $\{W_n\}$ and the fields from Eq (\ref{h_Edist}),  we solve  Eq (\ref{zstar}) numerically to give $z_*(\varphi )$.  If we substitute into Eq (\ref{Pphi}) we see that all terms except a factor $e^{-NU(\varphi)}$ are determined, so we can ``read off'' $U(\varphi )$  from the measured distribution of activity $P_{\rm exp}(\varphi )$ along the projection.

\begin{figure}[t]
\includegraphics[width=\linewidth]{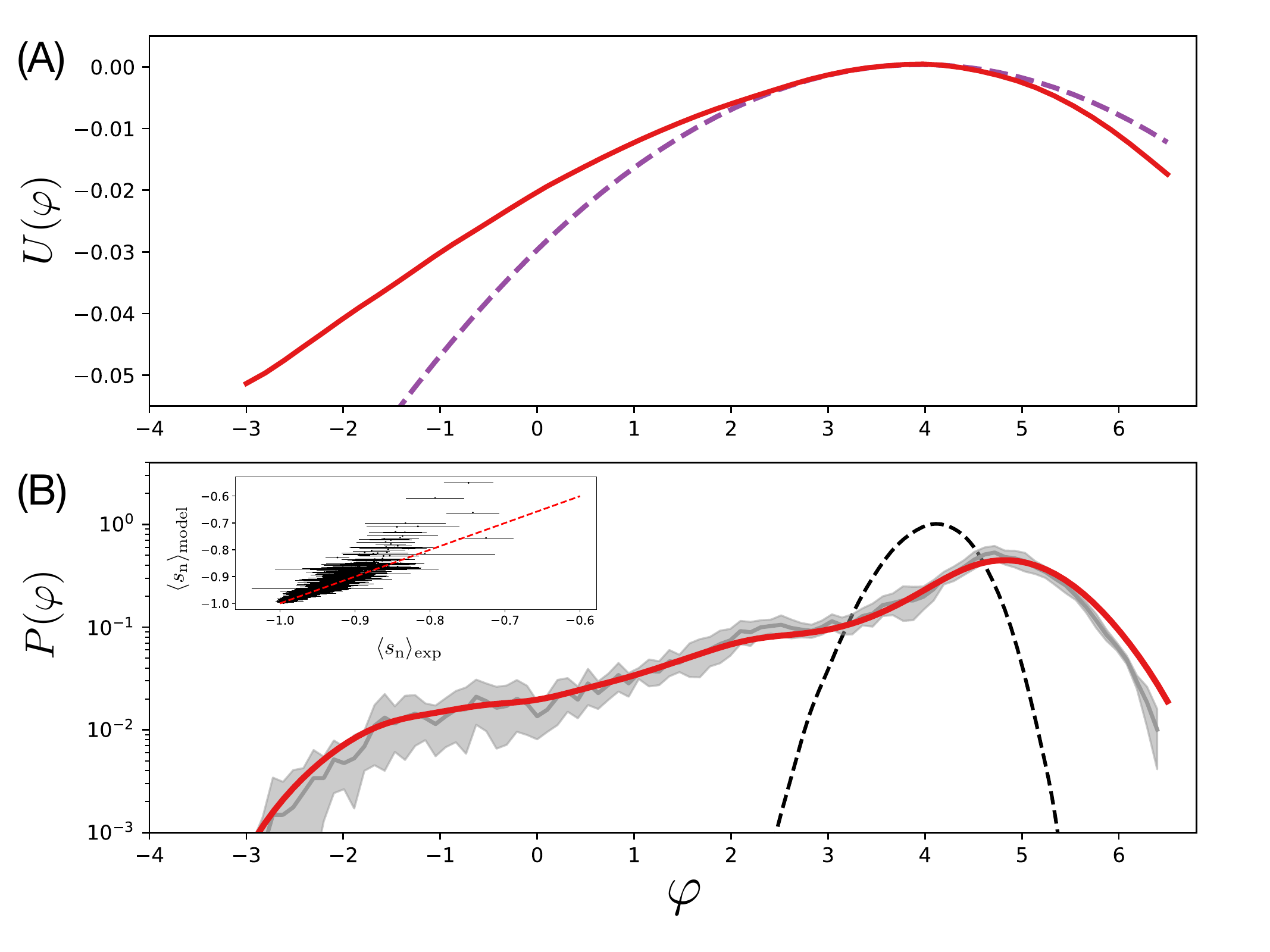}
\caption{Maximum entropy model that matches the mean activity of each neuron and the distribution of activity along a single projection, Eq~(\ref{Edist}), for $N=1416$ neurons in the mouse hippocampus.   The projection corresponds to the maximum eigenvalue of the correlation matrix.  (A) The potential $U(\varphi )$ (solid red) compared with a quadratic (black dashed). (B) The distribution of activity $P(\varphi )$  predicted by the model (red) and estimated from the data (grey; width shows standard deviation across tenths of the data); compare with the expected results for independent neurons (dashed). Inset shows the mean activity of each neuron, model vs data. \label{fig:dist}}
\end{figure}

We have applied this analysis to experiments on
% $N=1416$ neurons in the CA1 region of 
the mouse hippocampus, with results in Fig \ref{fig:dist}.  Following the guide above, we choose the projection corresponding to the largest eigenvalue of the correlation matrix, scaled as in   Eq (\ref{WWtilde}).  We see in Fig \ref{fig:dist}A that the potential $U(\varphi )$ departs significantly from quadratic at large $|\varphi|$, hinting as to why this model might ``work'' even when fixing the variance of activity along the projection fails, as above.    

%Our saddle point approximation is a generalized mean-field theory.  
%But we have seen above that mean-field theories fail when applied to real data because these data drive parameters of the model close to a first order phase transition.  To see what happens in this improved model 
To show that the data are in a regime where our generalized mean-field theory works
%this approximation works 
we performed a Monte Carlo simulation with energy from Eq (\ref{Edist}), the potential from Fig \ref{fig:dist}A, and fields from Eq (\ref{h_Edist}).  Figure \ref{fig:dist}B shows that we correctly reproduce the distribution of activity along the projection;
%which is far from what would be expected from independent neurons.  This suggests that we are effectively capturing the correlation structure in a generalized mean-field theory, and 
the inset shows that we recover the correct mean activities as well.  Note that these are not perfect reconstructions, but good approximations.  The troublesome local minima have been banished.

\begin{figure}[b]
    \centering
    \includegraphics[width=\linewidth]{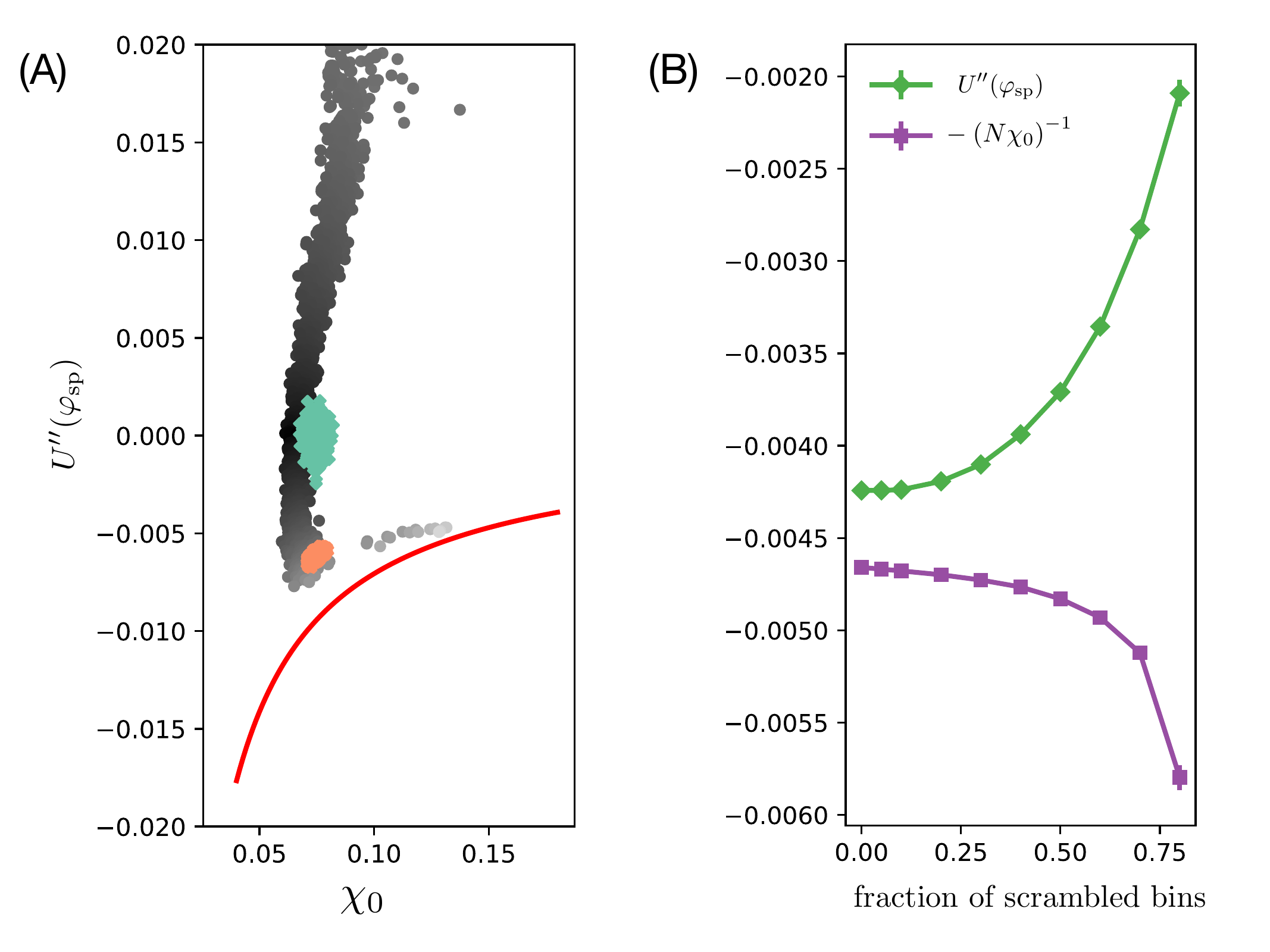}
    \caption{Approach to a critical point in matching the distribution of activity along a projection,  for $N=1416$ neurons in the mouse hippocampus.   (A) Scatter plot of $U''(\varphi_{\rm sp})$ vs $\chi_0$ for different projections: random (green), random positive (red), and the eigenvectors (grey) of the correlation matrix as in Eq (\ref{WWtilde}).  The red line marks the critical parameter settings where $NU''(\varphi_{\rm sp}) + \chi_0^{-1} = 0$.
 (B) Trajectories of $U''(\varphi_{\rm sp})$ and $-(N\chi_0)^{-1}$ as we look at networks with same mean activity for each neuron but weaker correlations, generated by shuffling a fraction of the time bins independently for each neuron.
    \label{fig:critical}}
\end{figure}

For some parameter values the curvature of the free energy will vanish at the saddle point, signaling the proximity of a second  order phase transition. 
%When we use a saddle point approximation to evaluate the partition function in Eq (\ref{Zdist3}), there is a danger that the curvature of the free energy vanishes at the saddle point; this signals the proximity of a second  order phase transition.  
The condition for this to happen is $NU''(\varphi_{\rm sp}) + \chi_0^{-1}=0$, where $\chi_0$ is the variance of activity along the projection that we would find if the neurons were independent,   
\begin{equation}
\chi_0 = \sum_{n=1}^N W_n^2\left[ 1 -\langle s_n\rangle^2\right] .
\end{equation}
%normalized as with $\chi$ in Eq (\ref{chi_def1}).  
With random projections  $NU''(\varphi_{\rm sp})$ is near zero, far from criticality (Fig \ref{fig:critical}A).  If we choose random positive coefficients then $NU''(\varphi_{\rm sp})$ becomes consistently negative but still not close to $-\chi_0^{-1}$.   The more interesting case is when we choose coefficients related to the eigenvectors of the correlation matrix, as in Eq (\ref{WWtilde}).    Again most  choices are far from criticality, but we see a collection of modes that trail toward the boundary $NU''(\varphi_{\rm sp}) + \chi_0^{-1}=0$.

The approach to criticality depends on the strength of correlations in the network.  We can imagine systems in which the mean activity of each neuron is the same, but the correlations between pairs of neurons are  weaker, and we can generate   such data by   shuffling a fraction of the time bins independently for each neuron.  For each shuffled data set we repeat the construction above,  and we find that  $NU''(\varphi_{\rm sp})$ and $-\chi_0^{-1}$ gradually move apart as we consider less correlated networks; this shows that   plausible populations of neurons are farther from criticality than the real network. This is consistent with other signatures of near--critical behavior \cite{Meshulam+Bialek_2024}, including the scaling of these same data under coarse--graining \cite{meshulam2019RG}. 

\begin{figure}[t]
\includegraphics[width=.45\textwidth]{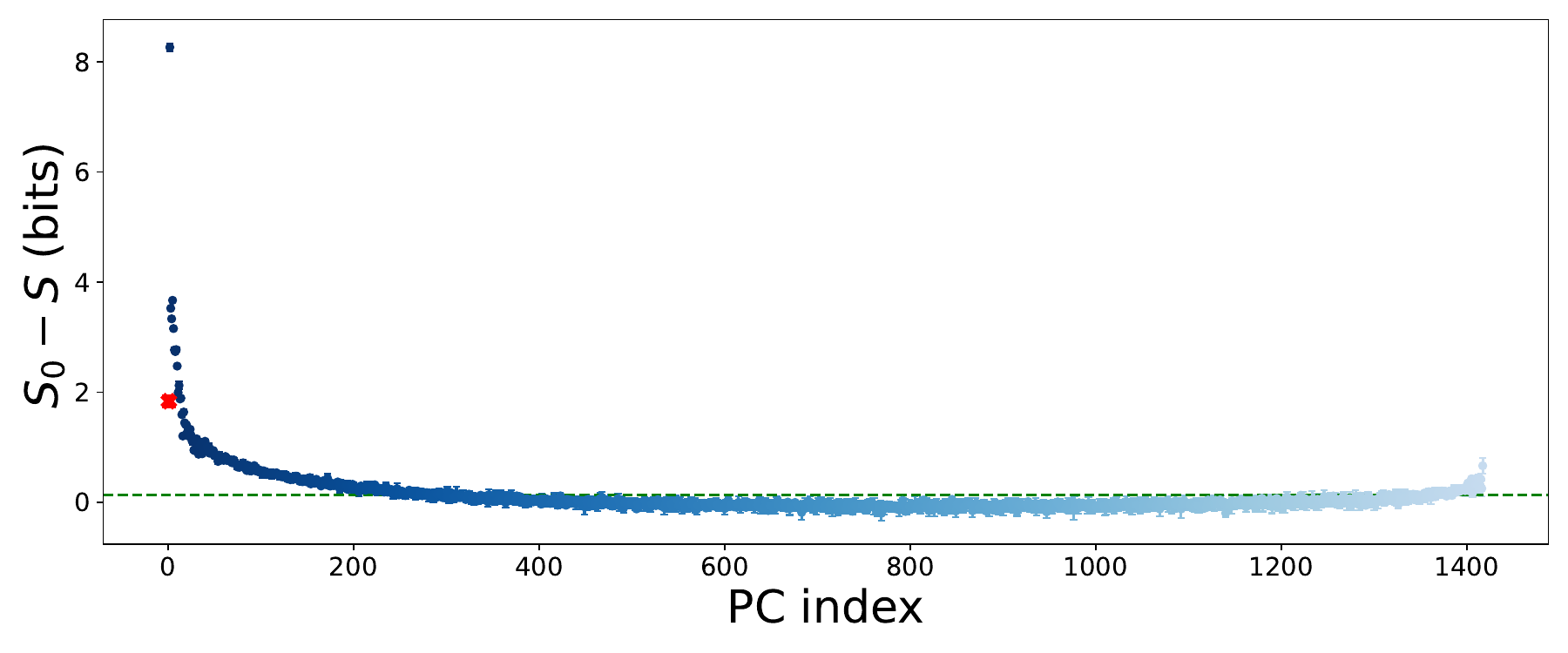}
\caption{\label{fig:entropy_projections}Difference between the entropy of the independent model $S_0$ and the entropy of the model $S$ (Eq~\eqref{Edist}) for different choices of projection. Blue dots are $W$ from Eq (\ref{WWtilde}), with darker colors at larger $\rho_\alpha$. Red cross is a projection onto summed activity, and the horizontal dashed line is the independent entropy per neuron $S_0/N\sim0.13$ bits. }
\end{figure}

We can compute the entropy of the model in Eq.~\eqref{Edist} following the mean-field approximation,
\begin{equation}
S = S_0 + N \left(\langle U (\varphi)\rangle - U(\varphi_{\rm sp})\right)-\frac 12 \ln\left[NU''(\varphi_{\rm sp})\chi_0+1\right],
\end{equation}
with results in Fig \ref{fig:entropy_projections} for the same data as in Fig \ref{fig:critical}.
Matching the distribution along just one projection leads to an entropy reduction of  $\sim 5\%$ of the independent entropy.  There are roughly $200$ projections that each provide a contribution above the independent entropy per neuron. 

To summarize, it proves surprisingly difficult to construct a consistent mean-field theory for the patterns of activity in a network of real neurons.  We finally succeed with a model that matches the mean activity of individual neurons and the {\em distribution} of activity along one projection.  But there are many projections along which constraining the distribution generates a substantial reduction in entropy.  This suggests  that an important future direction is to extend the mean-field approximation to a model matching the distribution of multiple projections.  This would give a framework for describing networks of $N$ neurons with $\mathcal{O} (N)$ constraints, making it possible to analyze  experiments at much larger $N$.

\begin{acknowledgements}
We thank our experimental colleagues MJ Berry II, CD Brody, JL Gauthier, O Marre, and DW Tank for providing a guide to the data.  Work  supported in part by the  National Science Foundation, through the Center for the Physics of Biological Function (PHY--1734030); and by fellowships from the Human Frontiers Science Program (FM),   the James S.~McDonnell Foundation (CWL), the John Simon Guggenheim Memorial Foundation (WB),  and the Simons Foundation (WB and FM).
\end{acknowledgements}

\bibliography{refs_MFT}

\begin{thebibliography}{30}
\expandafter\ifx\csname natexlab\endcsname\relax\def\natexlab#1{#1}\fi
\expandafter\ifx\csname bibnamefont\endcsname\relax
  \def\bibnamefont#1{#1}\fi
\expandafter\ifx\csname bibfnamefont\endcsname\relax
  \def\bibfnamefont#1{#1}\fi
\expandafter\ifx\csname citenamefont\endcsname\relax
  \def\citenamefont#1{#1}\fi
\expandafter\ifx\csname url\endcsname\relax
  \def\url#1{\texttt{#1}}\fi
\expandafter\ifx\csname urlprefix\endcsname\relax\def\urlprefix{URL }\fi
\providecommand{\bibinfo}[2]{#2}
\providecommand{\eprint}[2][]{\url{#2}}

\bibitem[{\citenamefont{Meshulam and Bialek}(2024)}]{Meshulam+Bialek_2024}
\bibinfo{author}{\bibfnamefont{L.}~\bibnamefont{Meshulam}} \bibnamefont{and}
  \bibinfo{author}{\bibfnamefont{W.}~\bibnamefont{Bialek}},
  \bibinfo{journal}{arXiv preprint arXiv:2409.00412}  (\bibinfo{year}{2024}).

\bibitem[{\citenamefont{Hopfield}(1982)}]{hopfield1982}
\bibinfo{author}{\bibfnamefont{J.~J.} \bibnamefont{Hopfield}},
  \bibinfo{journal}{Proceedings of the National Academy of Sciences (USA)}
  \textbf{\bibinfo{volume}{79}}, \bibinfo{pages}{2554} (\bibinfo{year}{1982}).

\bibitem[{\citenamefont{Ackley et~al.}(1985)\citenamefont{Ackley, Hinton, and
  Sejnowski}}]{ackley+al_85}
\bibinfo{author}{\bibfnamefont{D.~H.} \bibnamefont{Ackley}},
  \bibinfo{author}{\bibfnamefont{G.~E.} \bibnamefont{Hinton}},
  \bibnamefont{and} \bibinfo{author}{\bibfnamefont{T.~J.}
  \bibnamefont{Sejnowski}}, \bibinfo{journal}{Cognitive Science}
  \textbf{\bibinfo{volume}{9}}, \bibinfo{pages}{147} (\bibinfo{year}{1985}).

\bibitem[{\citenamefont{Hopfield and Tank}(1985)}]{hopfield+tank1985}
\bibinfo{author}{\bibfnamefont{J.~J.} \bibnamefont{Hopfield}} \bibnamefont{and}
  \bibinfo{author}{\bibfnamefont{D.~W.} \bibnamefont{Tank}},
  \bibinfo{journal}{Biological Cybernetics} \textbf{\bibinfo{volume}{52}},
  \bibinfo{pages}{141} (\bibinfo{year}{1985}).

\bibitem[{\citenamefont{Hopfield and Tank}(1986)}]{hopfield+tank1986}
\bibinfo{author}{\bibfnamefont{J.~J.} \bibnamefont{Hopfield}} \bibnamefont{and}
  \bibinfo{author}{\bibfnamefont{D.~W.} \bibnamefont{Tank}},
  \bibinfo{journal}{Science} \textbf{\bibinfo{volume}{233}},
  \bibinfo{pages}{625} (\bibinfo{year}{1986}).

\bibitem[{\citenamefont{Schneidman et~al.}(2006)\citenamefont{Schneidman,
  Berry~II, Segev, and Bialek}}]{schneidman+al_2006}
\bibinfo{author}{\bibfnamefont{E.}~\bibnamefont{Schneidman}},
  \bibinfo{author}{\bibfnamefont{M.~J.} \bibnamefont{Berry~II}},
  \bibinfo{author}{\bibfnamefont{R.}~\bibnamefont{Segev}}, \bibnamefont{and}
  \bibinfo{author}{\bibfnamefont{W.}~\bibnamefont{Bialek}},
  \bibinfo{journal}{Nature} \textbf{\bibinfo{volume}{440}},
  \bibinfo{pages}{1007} (\bibinfo{year}{2006}).

\bibitem[{\citenamefont{Jaynes}(1957)}]{jaynes1957information}
\bibinfo{author}{\bibfnamefont{E.~T.} \bibnamefont{Jaynes}},
  \bibinfo{journal}{Physical Review} \textbf{\bibinfo{volume}{106}},
  \bibinfo{pages}{620} (\bibinfo{year}{1957}).

\bibitem[{\citenamefont{Jaynes}(1982)}]{jaynes1982rationale}
\bibinfo{author}{\bibfnamefont{E.~T.} \bibnamefont{Jaynes}},
  \bibinfo{journal}{Proceedings of the IEEE} \textbf{\bibinfo{volume}{70}},
  \bibinfo{pages}{939} (\bibinfo{year}{1982}).

\bibitem[{\citenamefont{Bialek et~al.}(2012)\citenamefont{Bialek, Cavagna,
  Giardina, Mora, Silvestri, Viale, and Walczak}}]{bialek2012statistical}
\bibinfo{author}{\bibfnamefont{W.}~\bibnamefont{Bialek}},
  \bibinfo{author}{\bibfnamefont{A.}~\bibnamefont{Cavagna}},
  \bibinfo{author}{\bibfnamefont{I.}~\bibnamefont{Giardina}},
  \bibinfo{author}{\bibfnamefont{T.}~\bibnamefont{Mora}},
  \bibinfo{author}{\bibfnamefont{E.}~\bibnamefont{Silvestri}},
  \bibinfo{author}{\bibfnamefont{M.}~\bibnamefont{Viale}}, \bibnamefont{and}
  \bibinfo{author}{\bibfnamefont{A.~M.} \bibnamefont{Walczak}},
  \bibinfo{journal}{Proceedings of the National Academy of Sciences (USA)}
  \textbf{\bibinfo{volume}{109}}, \bibinfo{pages}{4786} (\bibinfo{year}{2012}).

\bibitem[{\citenamefont{Cavagna et~al.}(2015)\citenamefont{Cavagna,
  Del~Castello, Dey, Giardina, Melillo, Parisi, and Viale}}]{cavagna+al_2015}
\bibinfo{author}{\bibfnamefont{A.}~\bibnamefont{Cavagna}},
  \bibinfo{author}{\bibfnamefont{L.}~\bibnamefont{Del~Castello}},
  \bibinfo{author}{\bibfnamefont{S.}~\bibnamefont{Dey}},
  \bibinfo{author}{\bibfnamefont{I.}~\bibnamefont{Giardina}},
  \bibinfo{author}{\bibfnamefont{S.}~\bibnamefont{Melillo}},
  \bibinfo{author}{\bibfnamefont{L.}~\bibnamefont{Parisi}}, \bibnamefont{and}
  \bibinfo{author}{\bibfnamefont{M.}~\bibnamefont{Viale}},
  \bibinfo{journal}{Physical Review E} \textbf{\bibinfo{volume}{92}},
  \bibinfo{pages}{012705} (\bibinfo{year}{2015}).

\bibitem[{\citenamefont{Okun et~al.}(2012)\citenamefont{Okun, Yger, Marguet,
  Gerard-Mercier, Benucci, Katzner, Busse, Carandini, and
  Harris}}]{okun+al_2012}
\bibinfo{author}{\bibfnamefont{M.}~\bibnamefont{Okun}},
  \bibinfo{author}{\bibfnamefont{P.}~\bibnamefont{Yger}},
  \bibinfo{author}{\bibfnamefont{S.~L.} \bibnamefont{Marguet}},
  \bibinfo{author}{\bibfnamefont{F.}~\bibnamefont{Gerard-Mercier}},
  \bibinfo{author}{\bibfnamefont{A.}~\bibnamefont{Benucci}},
  \bibinfo{author}{\bibfnamefont{S.}~\bibnamefont{Katzner}},
  \bibinfo{author}{\bibfnamefont{L.}~\bibnamefont{Busse}},
  \bibinfo{author}{\bibfnamefont{M.}~\bibnamefont{Carandini}},
  \bibnamefont{and} \bibinfo{author}{\bibfnamefont{K.~D.}
  \bibnamefont{Harris}}, \bibinfo{journal}{Journal of Neuroscience}
  \textbf{\bibinfo{volume}{32}}, \bibinfo{pages}{17108} (\bibinfo{year}{2012}).

\bibitem[{\citenamefont{Tka\v{c}ik et~al.}(2013)\citenamefont{Tka\v{c}ik,
  Marre, Mora, Amodei, Berry~II, and Bialek}}]{tkacik+al_2013}
\bibinfo{author}{\bibfnamefont{G.}~\bibnamefont{Tka\v{c}ik}},
  \bibinfo{author}{\bibfnamefont{O.}~\bibnamefont{Marre}},
  \bibinfo{author}{\bibfnamefont{T.}~\bibnamefont{Mora}},
  \bibinfo{author}{\bibfnamefont{D.}~\bibnamefont{Amodei}},
  \bibinfo{author}{\bibfnamefont{M.~J.} \bibnamefont{Berry~II}},
  \bibnamefont{and} \bibinfo{author}{\bibfnamefont{W.}~\bibnamefont{Bialek}},
  \bibinfo{journal}{Journal of Statistical Mechanics Theory and Experiment}
  \textbf{\bibinfo{volume}{2013(3)}}, \bibinfo{pages}{P03011}
  (\bibinfo{year}{2013}).

\bibitem[{\citenamefont{Parisi}(1988)}]{parisi1988statistical}
\bibinfo{author}{\bibfnamefont{G.}~\bibnamefont{Parisi}},
  \emph{\bibinfo{title}{Statistical Field Theory}}
  (\bibinfo{publisher}{Frontiers in Physics, Addison-Wesley},
  \bibinfo{year}{1988}).

\bibitem[{\citenamefont{Sethna}(2021)}]{sethna2021statistical}
\bibinfo{author}{\bibfnamefont{J.}~\bibnamefont{Sethna}},
  \emph{\bibinfo{title}{Statistical Mechanics: Entropy, Order Parameters, and
  Complexity}} (\bibinfo{publisher}{Oxford University Press, Oxford},
  \bibinfo{year}{2021}).

\bibitem[{\citenamefont{Kivelson et~al.}(2024)\citenamefont{Kivelson, Jiang,
  and Chang}}]{kivelson+al_24}
\bibinfo{author}{\bibfnamefont{S.~A.} \bibnamefont{Kivelson}},
  \bibinfo{author}{\bibfnamefont{J.~M.} \bibnamefont{Jiang}}, \bibnamefont{and}
  \bibinfo{author}{\bibfnamefont{J.}~\bibnamefont{Chang}},
  \emph{\bibinfo{title}{Statistical Mechanics of Phases and Phase Transitions}}
  (\bibinfo{publisher}{Princeton University Press, Princeton},
  \bibinfo{year}{2024}).

\bibitem[{\citenamefont{Sahani}(1999)}]{sahani_99}
\bibinfo{author}{\bibfnamefont{M.}~\bibnamefont{Sahani}}, Ph.D. thesis,
  \bibinfo{school}{California Institute of Technology},
  \bibinfo{address}{Pasdena, CA} (\bibinfo{year}{1999}).

\bibitem[{\citenamefont{Whiteway and Butts}(2019)}]{whiteway+butts_2019}
\bibinfo{author}{\bibfnamefont{M.~R.} \bibnamefont{Whiteway}} \bibnamefont{and}
  \bibinfo{author}{\bibfnamefont{D.~A.} \bibnamefont{Butts}},
  \bibinfo{journal}{Current Opinion in Neurobiology}
  \textbf{\bibinfo{volume}{58}}, \bibinfo{pages}{86} (\bibinfo{year}{2019}).

\bibitem[{\citenamefont{Cunningham and Yu}(2014)}]{Cunningham2014}
\bibinfo{author}{\bibfnamefont{J.~P.} \bibnamefont{Cunningham}}
  \bibnamefont{and} \bibinfo{author}{\bibfnamefont{B.~M.} \bibnamefont{Yu}},
  \bibinfo{journal}{Nature Neuroscience} \textbf{\bibinfo{volume}{17}},
  \bibinfo{pages}{1500} (\bibinfo{year}{2014}).

\bibitem[{\citenamefont{Gallego et~al.}(2017)\citenamefont{Gallego, Perich,
  Miller, and Solla}}]{gallego2017neural}
\bibinfo{author}{\bibfnamefont{J.~A.} \bibnamefont{Gallego}},
  \bibinfo{author}{\bibfnamefont{M.~G.} \bibnamefont{Perich}},
  \bibinfo{author}{\bibfnamefont{L.~E.} \bibnamefont{Miller}},
  \bibnamefont{and} \bibinfo{author}{\bibfnamefont{S.~A.} \bibnamefont{Solla}},
  \bibinfo{journal}{Neuron} \textbf{\bibinfo{volume}{94}}, \bibinfo{pages}{978}
  (\bibinfo{year}{2017}).

\bibitem[{\citenamefont{Nieh et~al.}(2021)\citenamefont{Nieh, Schottdorf,
  Freeman, Low, Lewallen, Koay, Pinto, Gauthier, Brody, and Tank}}]{Edward2021}
\bibinfo{author}{\bibfnamefont{E.~H.} \bibnamefont{Nieh}},
  \bibinfo{author}{\bibfnamefont{M.}~\bibnamefont{Schottdorf}},
  \bibinfo{author}{\bibfnamefont{N.~W.} \bibnamefont{Freeman}},
  \bibinfo{author}{\bibfnamefont{R.~J.} \bibnamefont{Low}},
  \bibinfo{author}{\bibfnamefont{S.}~\bibnamefont{Lewallen}},
  \bibinfo{author}{\bibfnamefont{S.~A.} \bibnamefont{Koay}},
  \bibinfo{author}{\bibfnamefont{L.}~\bibnamefont{Pinto}},
  \bibinfo{author}{\bibfnamefont{J.~L.} \bibnamefont{Gauthier}},
  \bibinfo{author}{\bibfnamefont{C.~D.} \bibnamefont{Brody}}, \bibnamefont{and}
  \bibinfo{author}{\bibfnamefont{D.~W.} \bibnamefont{Tank}},
  \bibinfo{journal}{Nature} \textbf{\bibinfo{volume}{595}}, \bibinfo{pages}{80}
  (\bibinfo{year}{2021}).

\bibitem[{\citenamefont{Cocco et~al.}(2011)\citenamefont{Cocco, Monasson, and
  Sessak}}]{Cocco2011}
\bibinfo{author}{\bibfnamefont{S.}~\bibnamefont{Cocco}},
  \bibinfo{author}{\bibfnamefont{R.}~\bibnamefont{Monasson}}, \bibnamefont{and}
  \bibinfo{author}{\bibfnamefont{V.}~\bibnamefont{Sessak}},
  \bibinfo{journal}{Physical Review E - Statistical, Nonlinear, and Soft Matter
  Physics} \textbf{\bibinfo{volume}{83}} (\bibinfo{year}{2011}).

\bibitem[{\citenamefont{Amit et~al.}(1987)\citenamefont{Amit, Gutfreund, and
  Sompolinsky}}]{amit+al_1987}
\bibinfo{author}{\bibfnamefont{D.~J.} \bibnamefont{Amit}},
  \bibinfo{author}{\bibfnamefont{H.}~\bibnamefont{Gutfreund}},
  \bibnamefont{and}
  \bibinfo{author}{\bibfnamefont{H.}~\bibnamefont{Sompolinsky}},
  \bibinfo{journal}{Annals of Physics} \textbf{\bibinfo{volume}{173}},
  \bibinfo{pages}{30} (\bibinfo{year}{1987}).

\bibitem[{\citenamefont{Krotov and Hopfield}(2016)}]{krotov+hopfield_2016}
\bibinfo{author}{\bibfnamefont{D.}~\bibnamefont{Krotov}} \bibnamefont{and}
  \bibinfo{author}{\bibfnamefont{J.~J.} \bibnamefont{Hopfield}}, in
  \emph{\bibinfo{booktitle}{Advances in Neural Information Processing
  Systems}}, edited by \bibinfo{editor}{\bibfnamefont{D.}~\bibnamefont{Lee}},
  \bibinfo{editor}{\bibfnamefont{M.}~\bibnamefont{Sugiyama}},
  \bibinfo{editor}{\bibfnamefont{U.}~\bibnamefont{Luxburg}},
  \bibinfo{editor}{\bibfnamefont{I.}~\bibnamefont{Guyon}}, \bibnamefont{and}
  \bibinfo{editor}{\bibfnamefont{R.}~\bibnamefont{Garnett}}
  (\bibinfo{publisher}{Curran Associates, Inc.}, \bibinfo{year}{2016}),
  vol.~\bibinfo{volume}{29}, pp. \bibinfo{pages}{1172--1180}.

\bibitem[{\citenamefont{Di~Carlo et~al.}(2025)\citenamefont{Di~Carlo, Mignacco,
  Lynn, and Bialek}}]{dicarlo+al_2025}
\bibinfo{author}{\bibfnamefont{L.}~\bibnamefont{Di~Carlo}},
  \bibinfo{author}{\bibfnamefont{F.}~\bibnamefont{Mignacco}},
  \bibinfo{author}{\bibfnamefont{C.~W.} \bibnamefont{Lynn}}, \bibnamefont{and}
  \bibinfo{author}{\bibfnamefont{W.}~\bibnamefont{Bialek}}
  (\bibinfo{year}{2025}), \bibinfo{note}{in preparation}.

\bibitem[{\citenamefont{Tka{\v{c}}ik et~al.}(2014)\citenamefont{Tka{\v{c}}ik,
  Marre, Amodei, Schneidman, Bialek, and Berry}}]{tkavcik2014searching}
\bibinfo{author}{\bibfnamefont{G.}~\bibnamefont{Tka{\v{c}}ik}},
  \bibinfo{author}{\bibfnamefont{O.}~\bibnamefont{Marre}},
  \bibinfo{author}{\bibfnamefont{D.}~\bibnamefont{Amodei}},
  \bibinfo{author}{\bibfnamefont{E.}~\bibnamefont{Schneidman}},
  \bibinfo{author}{\bibfnamefont{W.}~\bibnamefont{Bialek}}, \bibnamefont{and}
  \bibinfo{author}{\bibfnamefont{M.~J.} \bibnamefont{Berry}},
  \bibinfo{journal}{PLoS Computational Biology} \textbf{\bibinfo{volume}{10}},
  \bibinfo{pages}{e1003408} (\bibinfo{year}{2014}).

\bibitem[{\citenamefont{{Allen Institute MindScope Program Allen Brain
  Observatory}}()}]{AllenData}
\bibinfo{author}{\bibnamefont{{Allen Institute MindScope Program Allen Brain
  Observatory}}}, \emph{\bibinfo{title}{Neuropixels visual coding (dataset)
  2019}}, \bibinfo{howpublished}{https://brain-map.org/explore/circuits}.

\bibitem[{\citenamefont{Gauthier and Tank}(2018)}]{gauthier2018dedicated}
\bibinfo{author}{\bibfnamefont{J.~L.} \bibnamefont{Gauthier}} \bibnamefont{and}
  \bibinfo{author}{\bibfnamefont{D.~W.} \bibnamefont{Tank}},
  \bibinfo{journal}{Neuron} \textbf{\bibinfo{volume}{99}}, \bibinfo{pages}{179}
  (\bibinfo{year}{2018}).

\bibitem[{\citenamefont{Meshulam et~al.}(2019)\citenamefont{Meshulam, Gauthier,
  Brody, Tank, and Bialek}}]{meshulam2019RG}
\bibinfo{author}{\bibfnamefont{L.}~\bibnamefont{Meshulam}},
  \bibinfo{author}{\bibfnamefont{J.~L.} \bibnamefont{Gauthier}},
  \bibinfo{author}{\bibfnamefont{C.~D.} \bibnamefont{Brody}},
  \bibinfo{author}{\bibfnamefont{D.~W.} \bibnamefont{Tank}}, \bibnamefont{and}
  \bibinfo{author}{\bibfnamefont{W.}~\bibnamefont{Bialek}},
  \bibinfo{journal}{Physical Review Letters} \textbf{\bibinfo{volume}{123}},
  \bibinfo{pages}{178103} (\bibinfo{year}{2019}).

\bibitem[{\citenamefont{Meshulam et~al.}(2021)\citenamefont{Meshulam, Gauthier,
  Brody, Tank, and Bialek}}]{Meshulam2021}
\bibinfo{author}{\bibfnamefont{L.}~\bibnamefont{Meshulam}},
  \bibinfo{author}{\bibfnamefont{J.~L.} \bibnamefont{Gauthier}},
  \bibinfo{author}{\bibfnamefont{C.~D.} \bibnamefont{Brody}},
  \bibinfo{author}{\bibfnamefont{D.~W.} \bibnamefont{Tank}}, \bibnamefont{and}
  \bibinfo{author}{\bibfnamefont{W.}~\bibnamefont{Bialek}},
  \bibinfo{journal}{arXiv preprint arXiv:2112.14735}  (\bibinfo{year}{2021}).

\bibitem[{\citenamefont{Lynn et~al.}(2023)\citenamefont{Lynn, Yu, Pang, Palmer,
  and Bialek}}]{Lynn2023}
\bibinfo{author}{\bibfnamefont{C.~W.} \bibnamefont{Lynn}},
  \bibinfo{author}{\bibfnamefont{Q.}~\bibnamefont{Yu}},
  \bibinfo{author}{\bibfnamefont{R.}~\bibnamefont{Pang}},
  \bibinfo{author}{\bibfnamefont{S.~E.} \bibnamefont{Palmer}},
  \bibnamefont{and} \bibinfo{author}{\bibfnamefont{W.}~\bibnamefont{Bialek}}
  (\bibinfo{year}{2023}), \urlprefix\url{http://arxiv.org/abs/2402.00007}.

\end{thebibliography}

\end{document}